\documentclass[times,twocolumn,final]{elsarticle}

\usepackage{arxiv}
\usepackage{graphicx}%
\usepackage{multirow}%
\usepackage{amsmath,amssymb,amsfonts}%

\usepackage{amsthm}%
\usepackage{mathrsfs}%
\usepackage[title]{appendix}%
\usepackage{xcolor}%
\usepackage{booktabs}%
\usepackage{algpseudocode,algorithm}
\MakeRobust{\Call}
\usepackage{pifont}
\usepackage{adjustbox}
\usepackage{graphicx}
\usepackage{float}
\usepackage{placeins}
\usepackage{fancyhdr}
\DeclareUnicodeCharacter{2032}{\ensuremath{'} }
\UseRawInputEncoding
\usepackage[hyphens]{url}
\usepackage{hyperref}

\fancypagestyle{firstpagestyle}{
    \fancyhf{} 
    \fancyhead{} 
    \fancyfoot{} 
}

\fancypagestyle{default}{
    \fancyhf{}
    \fancyhead[R]{\thepage} 
}

\begin{document}

\title{Surgeons’ Awareness, Expectations, and Involvement with Artificial Intelligence: a Survey Pre and Post the GPT Era}

\author[1]{Lorenzo \snm{Arboit}\corref{corresp}}
\cortext[corresp]{Corresponding author: \texttt{larboit@unistra.fr}}
\author[1]{Dennis N. \snm{Schneider}}
\author[2]{Toby \snm{Collins}}
\author[3]{Daniel A. \snm{Hashimoto}}
\author[2,4,5]{Silvana \snm{Perretta}}
\author[5]{Bernard \snm{Dallemagne}}
\author[2]{Jacques \snm{Marescaux}}
\author{EAES Working Group}
\author[1,4]{Nicolas \snm{Padoy}\fnref{equal}}
\author[4,6]{Pietro \snm{Mascagni}\fnref{equal}}

\address[1]{University of Strasbourg, CNRS, INSERM, ICube, UMR7357, Strasbourg, France}
\address[2]{IRCAD, Strasbourg, France}
\address[3]{University of Pennsylvania Perelman School of Medicine, Department of Surgery, Philadelphia, PA, USA}
\address[4]{Institute of Image-Guided Surgery, Strasbourg, France}
\address[5]{The University Hospitals of Strasbourg, Strasbourg, France}
\address[6]{Fondazione Policlinico Universitario Agostino Gemelli IRCCS, Rome, Italy}

\fntext[equal]{Co-last authors.}

\received{XXX}
\finalform{XXX}
\accepted{XXX}
\availableonline{XXX}
\communicated{XXX}

\begin{abstract}

\textbf{Introduction:} Artificial Intelligence (AI) is transforming medicine, with generative AI models like ChatGPT reshaping perceptions of its potential. This study examines surgeons’ awareness, expectations, and involvement with AI in surgery through comparative surveys conducted in 2021 and 2024.

\textbf{Methods:} Two cross-sectional surveys were distributed globally in 2021 and 2024, the first before an IRCAD webinar and the second during the annual EAES meeting. The surveys assessed demographics, AI awareness, expectations, involvement, and ethics (2024 only).

\textbf{Results:} The surveys collected a total of 671 responses from 98 countries, 522 and 149 in 2021 and 2024, respectively. Awareness of AI courses rose from 14.5\% in 2021 to 44.6\% in 2024, while course attendance increased from 12.9\% to 23\%. Despite this, familiarity with foundational AI concepts remained limited. Expectations for AI’s role shifted in 2024, with hospital management gaining relevance. Ethical concerns gained prominence, with 87.2\% of 2024 participants emphasizing accountability and transparency. Infrastructure limitations remained the primary obstacle to the implementation of AI. Interdisciplinary collaboration and structured training were identified as critical for successful AI adoption. Optimism about AI’s transformative potential remained high, with 79.9\% of respondents believing AI would positively impact surgery and 96.6\% of surgeons willing to integrate AI into their clinical practice.

\textbf{Conclusion:} Surgeons’ perceptions of AI are evolving, driven by the rise of generative AI and the advancements in surgical data science. While enthusiasm for integration is strong, knowledge gaps and infrastructural challenges persist. Addressing these through education, ethical frameworks, and infrastructure development is essential.

\end{abstract}

\maketitle
\thispagestyle{firstpagestyle}

\section{Introduction}
\label{sec:introduction}

Artificial Intelligence (AI) in medicine and surgery is often surrounded by a mix of inflated expectations or skepticism~\cite{1}. This is particularly true in recent years, with the rise of generative AI (GenAI) models such as ChatGPT (OpenAI, San Francisco, CA, USA) and Midjourney (Midjourney, Inc., San Francisco, CA, USA), which have brought AI into the public eye~\cite{2}. These widely available platforms have not only showcased AI’s potential but have also raised new concerns, especially within the medical community~\cite{3}. Although AI has yet to impact surgical practice significantly~\cite{4}, the conversation around its integration is intensifying. Organizations like the European Association for Endoscopic Surgery (EAES) and other surgical bodies have recognized this potential, actively promoting educational events to introduce AI research and its clinical applications to surgeons.

Generalist AI and Surgical Data Science (SDS) are two fast-evolving fields~\cite{5,6}. Generalist AI, referring to AI models not developed specifically for surgery including large language models (LLMs) like ChatGPT and LLaMA~\cite{7}, has not yet been extensively applied in surgery~\cite{8} but it has already impacted medical research, with numerous reports showing its utility in areas such as trial design~\cite{9}, data analysis ~\cite{10,11}, and knowledge synthesis~\cite{12,13}. SDS, on the contrary, focuses on extracting, analyzing, and applying actionable insights from surgical data to improve decision-making, enhance outcomes, and optimize workflows. The field encompasses technologies such as computer vision, real-time data monitoring, and predictive analytics tailored specifically for the surgical domain~\cite{14}. Although still in its developmental stages, SDS has shown promising applications in areas like workflow analysis ~\cite{15,16,17,18}, skill assessment~\cite{15,19,20,21}, and error detection~\cite{20,22}, with increasing integration into clinical practice and operating room systems~\cite{23}.

Despite the surge of LLMs, GenAI represents only a small fraction of AI’s potential in medicine. Several advancements actually applied to the medical field come from computer vision applications, especially in diagnostics~\cite{24}. For instance, applications in radiology have grown rapidly, with several AI models approved for clinical use~\cite{25}. Similarly, AI-assisted screening colonoscopy has become the standard of care, with several randomized controlled trials showing a significant increase in adenoma detection rates thanks to computer-aided detection (CAD)~\cite{26,27}. In contrast, intraoperative decision support remains underdeveloped~\cite{14,28}, with only a handful of AI-enabled medical devices approved for clinical use~\cite{29}. Nonetheless, within the research community, there is growing interest in using AI to improve surgical performance by analyzing and reducing errors~\cite{30}, as well as enhancing surgical training programs~\cite{31}. The rapid implementations in these fields, however, face ethical concerns that many countries and institutions are now beginning to address. The European Union, for instance, has introduced the AI Act, establishing legal frameworks to regulate AI’s use in healthcare amongst others~\cite{32}.

This study aims to assess surgeons’ awareness, expectations, and involvement with AI in surgery, before and after the rise of mainstream generative AI applications like ChatGPT. By comparing data from surveys conducted in 2021 and 2024, this study seeks to identify key trends and changes in perception. We hope that the expectations and needs of surgeons, as captured through these surveys and now publicly shared, will guide research and development efforts to maximize the value of AI in surgery.

\section{Methods}
\label{sec:methods}

This study was reported according to the Checklist for Reporting OF Survey Studies (CROSS)~\cite{33}. The local institutional review board (IRB) waived the need for approval of the present study. No personally identifiable information was collected except for the email addresses of respondents, which were used solely to share survey results with participants who opted in.

\subsection{Survey}
\label{subsec:survey}

This study distributed two similarly structured cross-sectional, population-based online surveys in 2021 and 2024, and compared findings. The surveys were designed to assess surgeons’ awareness, expectations, and involvement with surgical AI both before and after the rise of mainstream generative AI models. Both surveys comprised the following sections: Demographics, AI Awareness, Expectations, Involvement, and Have Your Say. The 2024 survey incorporated an additional section, Ethical Considerations. Questions with Likert-like answers evaluating the importance of a specific topic were structured with 1 representing the highest level of importance. Table~\ref{table:survey_breakdown} provides a detailed breakdown of the number and distribution of questions across the different sections. Each survey was pretested: The first edition (2021) by the faculty of the IRCAD webinar ‘AI \& Surgery,’ while the second edition (2024) by members of the EAES Technology Committee. The complete questionnaires are available in \texttt{supplementary\_1.pdf} (2021 survey) and \texttt{supplementary\_2.pdf} (2024 survey).

\begin{table*}[t]
\caption{Surveys breakdown. \\ The number of questions per section is reported, with the number in parentheses corresponding to the number of optional free-text questions.}
\label{table:survey_breakdown}
\centering
\setlength{\tabcolsep}{10pt}
\begin{tabular}{lcc}
\\[0.5ex]
\textbf{Section Name} & \textbf{N. Questions - 2021} & \textbf{N. Questions - 2024} \\ \hline
Demographics & 10 & 10 \\
AI Awareness & 9 (2) & 12 \\
Expectations & 8 & 10 (1) \\
Involvement & 7 & 10 \\
Ethical Considerations & 0 & 11 \\
Have your say & 1 (1) & 1 (1) \\
\textbf{Total number of questions} & \textbf{35} & \textbf{54} \\
\end{tabular}
\end{table*}

\subsection{Population}
\label{subsec:population}

The target population for both surveys was surgeons, with no limitations regarding nationality, education level, age, specialty, hospital setting, or clinical experience. The surveys were written in English. The 2021 survey was shared through the IRCAD and WebSurg mailing list before the IRCAD ‘AI \& Surgery’ webinar, while the 2024 survey was disseminated during the EAES annual meeting via ad hoc notifications on the conference smartphone application. Both surveys were promoted through social media and institutional platforms, including LinkedIn and WebSurg. The 2021 survey collected responses from April to November 2021, while the 2024 survey was open from June to September 2024. 

\subsection{Statistical Analysis}
\label{subsec:analysis}

Data was collected using Google Forms (Google, Mountain View, CA, USA) and the Google Sheets plugin. A dedicated Google Colaboratory (Colab) notebook was developed to perform the primary data analysis. Responses from non-medical doctors were excluded. For questions originally using a Likert scale with one as the maximum score, the scale was inverted (1 = minimum score) to make the analysis and plots easier to interpret. Qualitative responses were processed using GPT-4o (OpenAI, San Francisco, CA, USA) for initial codification, followed by manual review and refinement based on established coding methodologies~\cite{34}, with answers categorized into specific groups based on content and assigned to all relevant groups if fitting multiple categories. A structured prompt for LLM analysis was based on the framework presented by Wachinger et al.~\cite{35}. Descriptive statistics were calculated to summarize the demographic characteristics of the participants and their responses, including measures of central tendency and variability for continuous variables and frequency distributions for categorical variables. A category-based chi-square test was used to analyze the distribution differences of demographic variables (excluding age) between groups, while ordinal variables, including age, were analyzed using the Mann-Whitney U test since normality testing showed a non-normal distribution. Post-hoc tests were performed on specific categories to compare the proportions for each category between the two surveys using Fisher’s exact tests with Bonferroni correction. Statistical significance was defined as p $<$ 0.05 for all analyses. All analyses were conducted using Colab and RStudio (Posit Software PBC, Boston, MA, USA), with statistical tests performed using Colab and R, version 4.4.1 (The R Foundation, Vienna, Austria), and plots generated within the Colab notebook and RStudio.

\section{Results}
\label{sec:results}

The data and the notebook are freely accessible at the following \href{https://colab.research.google.com/drive/1kQwmi4mdI1W0TYSk-EcdYc3CzNtcgSQ4?usp=sharing}{link} to ensure transparency, allow for reproducibility, and extension of the analysis.

\subsection{Demographics}
\label{subsec:demographics}

\begin{figure*}[h]
  \centering
  \includegraphics[width=\textwidth, trim=0cm 1.2cm 0cm 1.2cm, clip]{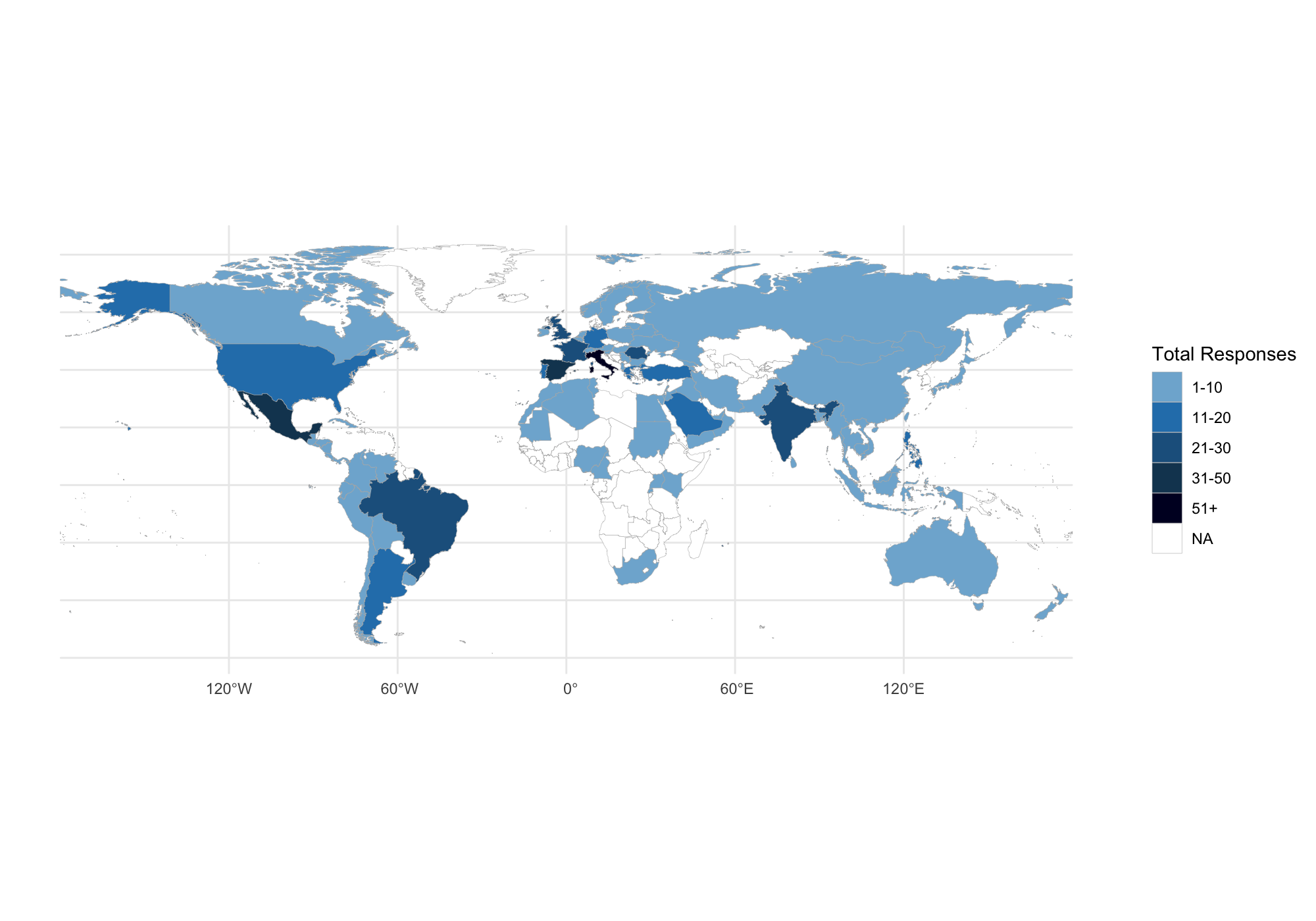}
  \caption{Global distribution of responders to the surgical AI surveys (2021 and 2024). \\ Distribution of included responders per country over the 2021 and 2024 editions.}
  \label{fig:global_distribution}
\end{figure*}

A total of 671 participants from 98 countries responded to the two surveys. Excluding non-doctors or non-medical students, 496 doctors from 86 countries responded to the 2021 survey, and 148 respondents from 55 countries for the 2024 edition. Figure~\ref{fig:global_distribution} illustrates the distribution of participants by country, with 47.1\% of responders based in Europe, and Italy representing 17.4\% of participants.

Respondents were predominantly male (80.6\% overall) and had significant clinical and research experience. Most were attending surgeons with over 10 years of experience (342 participants, 53.1\%), while 19.9\% were still in training (medical students, residents, and fellows). General surgeons comprised the majority (444 surgeons across 2021 and 2024, 68.9\% of the participants), followed by surgeons specialized in gynecology (8.9\%). Almost half worked in public hospitals (48.6\% of the total clinicians), and a large portion were regularly involved in research (45.3\% overall). Notably, participants in 2024 showed a significant difference only in scientific experience, specifically among those who were previously involved but are no longer doing research (13.9\% in 2021 vs 27.7\% in 2024, p = 0.001). Detailed demographics are provided in Table~\ref{table:demographics}.

\begin{table*}[t]
\caption{Demographics. Next to each category, p-values are reported to test the difference between the distribution of 2021 and 2024.}
\label{table:demographics}
\centering
\setlength{\tabcolsep}{8pt}
\begin{tabular}{lll}
\\[0.5ex]
\textbf{Item} & \textbf{2021 Number of responses (\%)} (total 496) & \textbf{2024 Number of responses (\%)} (total 148) \\[0.5ex] \hline

\multicolumn{3}{c}{\textit{Gender (p = 0.402)}} \\[0.5ex]
Male & 404 (81.4\%) & 115 (77.7\%) \\
Female & 90 (18.0\%) & 33 (22.3\%) \\
Prefer not to say & 2 (0.6\%) & 0 (0.0\%) \\[0.5ex]

\multicolumn{3}{c}{\textit{Age (p = 0.524)}} \\[0.5ex]
Mean & 45.1 & 45.7 \\
Max & 78 & 72 \\
Min & 22 & 24 \\[0.5ex]

\multicolumn{3}{c}{\textit{Specialty (p = 0.237)}} \\[0.5ex]
General Surgery & 293 (59.1\%) & 100 (67.6\%) \\
Colon and Rectal Surgery & 39 (7.8\%) & 12 (8.1\%) \\
Obstetrics and Gynecology & 46 (9.3\%) & 11 (7.4\%) \\
Other & 118 (23.8\%) & 25 (16.9\%) \\[0.5ex]

\multicolumn{3}{c}{\textit{Main Hospital Setting (p = 0.302)}} \\[0.5ex]
Private Practice & 102 (20.6\%) & 38 (25.7\%) \\
Public Hospital & 241 (48.6\%) & 73 (48.6\%) \\
Academic Center & 153 (30.8\%) & 38 (25.7\%) \\[0.5ex]

\multicolumn{3}{c}{\textit{Clinical Experience (p = 0.837)}} \\[0.5ex]
Medical Student & 13 (2.6\%) & 6 (4.1\%) \\
Resident & 65 (13.1\%) & 17 (11.5\%) \\
Fellow & 21 (4.2\%) & 6 (4.1\%) \\
Consultant/Attending $\leq$ 10 years & 131 (26.4\%) & 43 (29.0\%) \\
Consultant/Attending $>$ 10 years & 266 (53.7\%) & 76 (51.3\%) \\[0.5ex]

\multicolumn{3}{c}{\textit{Scientific Experience (p = 0.002)\textsuperscript{a}}} \\[0.5ex]
None & 40 (8.1\%) & 11 (7.4\%) \\
Involved, not anymore & 69 (13.9\%) & 41 (27.7\%) \\
Occasionally involved & 142 (28.6\%) & 35 (23.7\%) \\
Regularly involved & 117 (23.6\%) & 27 (18.2\%) \\
Research fellow / PhD Candidate & 38 (7.7\%) & 16 (10.8\%) \\
Research projects leader & 90 (18.1\%) & 18 (12.2\%) \\[0.5ex]

\multicolumn{3}{c}{\textit{Research Experience (p = 0.236)}} \\[0.5ex]
No research experience & 44 (8.9\%) & 20 (13.5\%) \\
$<$1 year & 48 (9.7\%) & 8 (5.4\%) \\
1–3 years & 113 (22.8\%) & 29 (19.6\%) \\
3–10 years & 132 (26.6\%) & 41 (27.7\%) \\
$>$10 years & 159 (32.0\%) & 50 (33.8\%) \\[0.5ex]
\end{tabular}

\vspace{0.5em}
\raggedright
\textsuperscript{a}Post-hoc analysis for statistically significant categories (scientific experience) revealed that only the distribution of surgeons who were involved in scientific research but not anymore changed ($p = 0.001$).
\end{table*}

\subsection{AI Awareness}
\label{subsec:awareness}

\begin{figure*}[t]
  \centering
  \includegraphics[width=\textwidth, trim=2cm 0cm 2cm 0cm, clip]{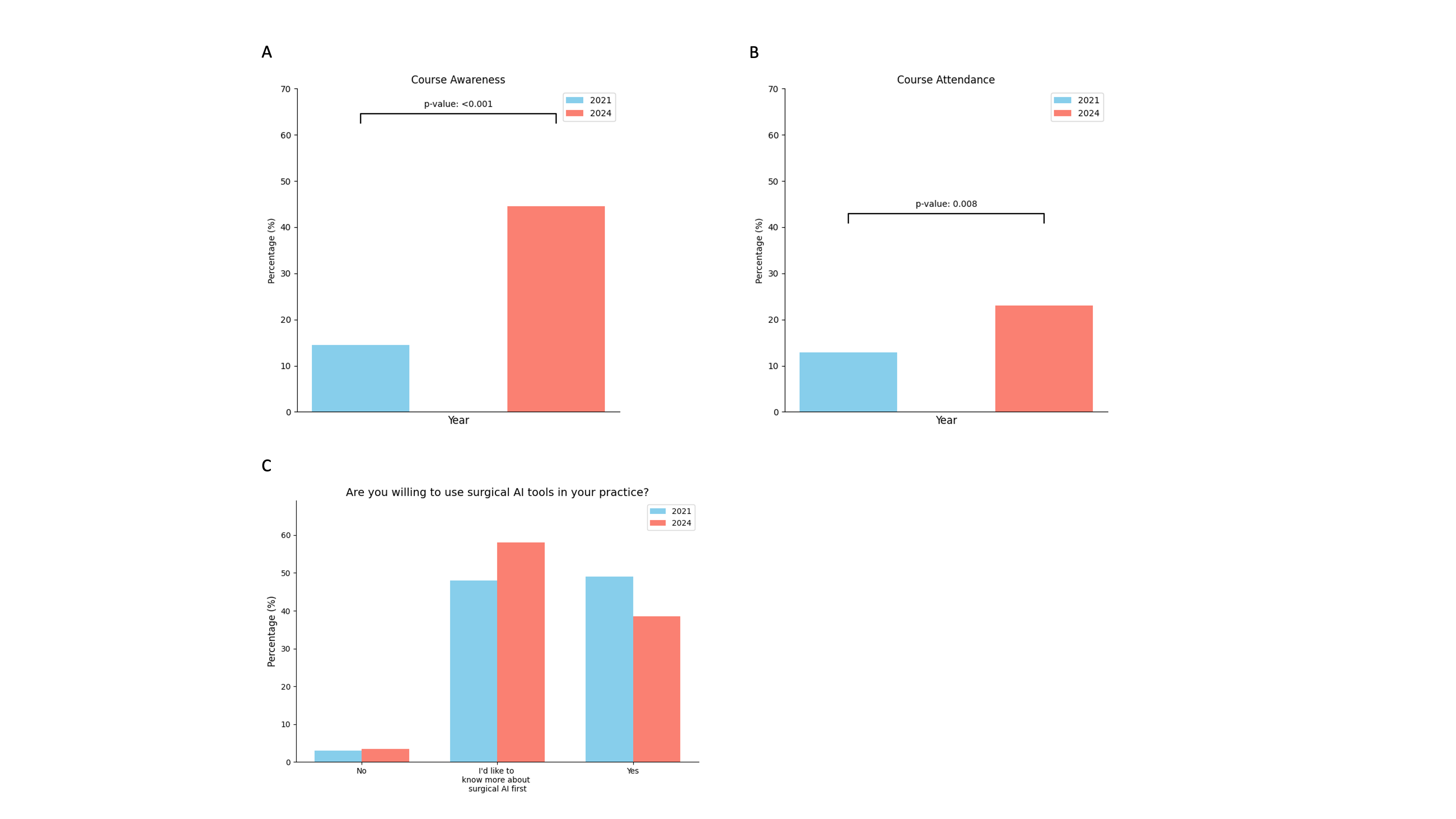}
  \caption{Surgeons’ awareness, course attendance, and willingness to adopt AI in clinical practice (2021 vs 2024). \\ Significant comparisons in the sections “AI Awareness” and “Involvement”. A) Changes in the awareness of surgeons in AI-oriented courses. B) Changes in the attendance of surgeons in AI-oriented courses. C) Responders’ willingness to use AI within their clinical practice. The statistical significance of the comparison between the response distributions was determined to be p = 0.030. However, no significant difference was observed when comparing individual responses.}
  \label{fig:results}
\end{figure*}

On average, 73.0\% of participants struggled with foundational AI concepts, such as machine learning, deep learning, and computer vision. Familiarity with these terms and programming skills remained static between 2021 and 2024, as indicated by similar proportions in key categories (e.g., moderate familiarity: 36.5\% in 2021 vs. 25.7\% in 2024, p = 0.087). However, awareness of AI courses significantly increased by 2024 (44.6\% vs. 14.5\%, p $<$ 0.001) as shown in Figure~\ref{fig:results}A, with platforms like Coursera, Udemy, and offerings from surgical associations suggested as key resources. More clinicians attended AI courses in 2024 (23.0\% vs. 12.9\%, p = 0.008) (Figure~\ref{fig:results}B), favoring online formats. Interest in AI education remained high (92.5\% in 2021 vs 91.2\% in 2024, p = 1.000), with 44.7\% supporting its integration into medical and surgical training. By 2024, 72.4\% of participants had used GenAI tools, with 60.5\% reporting usage of ChatGPT specifically.

\subsection{AI Expectations}
\label{subsec:expectations}

Respondents across both years believed AI would be integrated into surgery, with an average of 94.8\% expressing this belief across the two surveys. However, the timeframe for integration shortened significantly between the surveys, with the proportion of respondents indicating a timeframe of less than 5 years increasing from 32.1\% in 2021 to 55.4\% in 2024 (p $<$ 0.001). 

In 2021, participants expected AI to have the greatest impact on preoperative (mean = 3.15, on a scale of 1 to 4) and intraoperative (mean = 2.93) stages of surgical care, followed by hospital management (mean = 2.85) and postoperative care (mean = 2.56). By 2024, expectations shifted, with hospital management (mean = 3.25, p $<$ 0.001) and preoperative care (mean = 3.29) becoming the most anticipated areas of impact, while intraoperative care was deprioritized (mean = 2.64, p = 0.006); these results are reported in Figure~\ref{fig:barriers}A.

Surgeons highlighted specific areas in which they would expect to see AI play a more prominent role in ten years. Administrative roles gained prominence in 2024, significantly improving their perceived value (mean = 4.30 vs. 4.85 on a scale of 1 to 6, p $<$ 0.001). Enhanced surgical vision (mean = 4.88 vs. 5.12, p = 0.036) and AI’s role in training and education (mean = 4.88 vs. 5.19, p = 0.045) also saw increased recognition as expected applications. These trends are detailed in Figure~\ref{fig:barriers}B.

Regarding the applications surgeons are interested in seeing implemented, there was a notable increase in interest in AI for administrative tasks (mean = 3.98 vs. 4.61, p $<$ 0.001) and surgical vision (mean = 4.92 vs. 5.19, p = 0.026). Conversely, interest in surgical automation declined (mean = 3.75 vs. 3.35, p = 0.021), as illustrated in Figure~\ref{fig:barriers}C.

In 2021, respondents with more scientific experience (correlation coefficient (c) = 0.0789, p = 0.041) and a better understanding of AI-related papers (c = 0.1165, p = 0.042) predicted an earlier integration of AI into surgical practice. Conversely, those who self-reported studying AI (c = -0.1921, p = 0.002) predicted a longer integration. By 2024, only technological affinity (c = 0.2568, p = 0.028) remained a predictor of an earlier integration.

Open-ended responses (117 in 2021; 35 in 2024) highlighted AI’s potential to improve perioperative care, reduce risks, and optimize workflows, alongside concerns about ethics and training needs. In 2021, there was specific interest in augmented reality (AR) and virtual reality (VR) for enhancing surgical vision, while in 2024, responses emphasized the importance of rigorous validation and AI serving as a supportive tool under surgeon control.

\subsection{Barriers to AI Adoption}
\label{subsec:barriers}

\begin{figure*}[t]
  \centering
  \includegraphics[width=\textwidth, trim=2cm 0cm 2cm 0cm, clip]{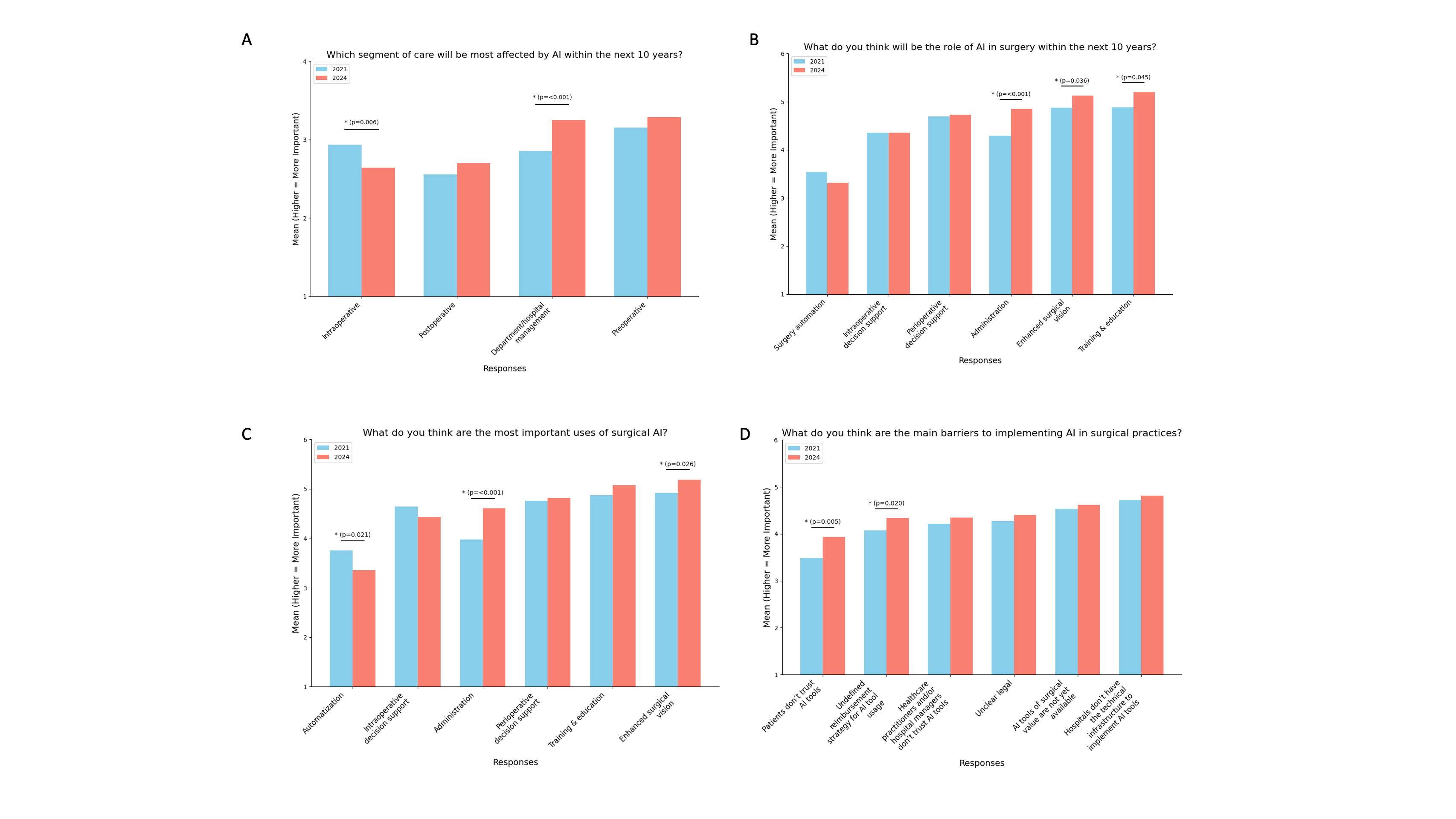}
  \caption{Comparative analysis of surgeons’ perspectives on AI’s impact, future applications, uses, and barriers in surgical practice (2021 vs 2024). \\ Distribution of mean responses and corresponding p-values from the Mann-Whitney U test for questions evaluated on a Likert scale. For question A, a 4-point Likert scale was used, while questions B, C, and D employed a 6-point Likert scale.}
  \label{fig:barriers}
\end{figure*}

In 2021 and 2024, responders considered the lack of technical infrastructure (average mean = 4.77 on a scale of 1 to 6) and the unavailability of surgical AI tools (average mean = 4.58) the primary barriers to AI adoption. Trust in AI tools by patients remained the least concerning barrier in both years but was significantly higher in 2024 (mean = 3.49 vs. 3.94, p = 0.005). Similarly, undefined reimbursement strategies were cited as the second to last barrier but concerns increased in the second edition (mean = 4.08 vs. 4.33, p = 0.020), as represented in Figure~\ref{fig:barriers}D. The 2024 survey also highlighted difficulties in video recording due to insufficient data storage (30.9\% of respondents) and recording devices (37.1\%), as reported in an ad-hoc question introduced in the second edition of the survey.

\subsection{Perception and Risks}
\label{subsec:perception}

An overall 79.9\% of respondents believed AI would positively impact surgery. Concerns about job displacement remained minimal ($\sim$14\% in both years). Respondents consistently prioritized accuracy over explainability (61.7\% in 2021 vs. 58.1\% in 2024, p = 1.000) when presented with a clinical scenario involving two AI tools for tumor characterization: one tool, which was more accurate in distinguishing between benign and malignant lesions, and another, which was less accurate but provided an explanation for its classification. Furthermore, in both years, familiarity with AI correlated with a preference for explainability over accuracy in AI tools, as demonstrated by participants familiar with AI-related terms in 2021 (c = 0.1181, p = 0.044) and surgeons who understood AI-focused articles in 2024 (c = 0.3403, p = 0.031). These findings are summarized in \texttt{supplementary\_3.pdf}. In 2024, qualitative answers showed that reduced human involvement was the most cited risk (14 mentions), followed by technical failures (10 mentions), data quality issues (9 mentions), and legal and ethical concerns (5 mentions each).

\subsection{Involvement}
\label{subsec:involvement}

On average, 20.5\% of surgeons were actively involved in AI research, while 33.9\% expressed interest in starting. In 2021, involvement in AI research was positively correlated with familiarity with AI terminology (r = 0.1127, p = 0.006) and scientific experience (r = 0.0956, p = 0.006). However, in 2024, experience with AI studies (r = 0.3926, p = 0.025) and collaborations with AI specialists (r = 0.4530, p $<$ 0.001) became positive predictors of familiarity with AI terminology. 

Over 80\% systematically recorded surgical videos, half of which were within approved research protocols. Additionally, an average of 17.1\% collected structured clinical and radiological datasets. Although collaboration with AI specialists was limited, interest in interdisciplinary efforts was substantial (69.4\% vs. 73.0\%, p = 1.000). Awareness of the SDS community was low in both editions. In 2021, 55.2\% of respondents reported not knowing about it, and in 2024, this figure increased to 61.5\% (p = 1.000). Importantly, an overwhelming average of 96.6\% of surgeons expressed a willingness to integrate AI into their clinical practice. Furthermore, more surgeons in 2024 showed interest in learning about AI before its practical implementation (48.0\% vs. 58.1\%, p = 0.095), as shown in Figure~\ref{fig:results}C; the comparison between the response distribution in this topic was statistically significant (p = 0.030). Willingness to use AI tools correlated with confidence in its integration (correlation coefficient (c)= 0.0916, p = 0.033) and previous AI studies (c = 0.0932, p = 0.017) in 2021 but not in 2024.

\subsection{Ethical Considerations}
\label{subsec:ethical}

Ethical considerations were only included in the survey in 2024 because they were not yet perceived as a significant issue in 2021, given the limited mainstream adoption of AI at that time. These were regarded as important by 87.2\% of respondents, though only 52.5\% expressed significant concern about potential ethical issues. Accountability and transparency were the top concerns (62.7\%), while the surgeon-patient relationship was considered least impacted (39.4\%). Multidisciplinary collaboration and comprehensive training were emphasized as key to ensuring responsible AI use and minimizing errors in AI-assisted treatment decisions. The full distribution of responses is reported in eTable 2 in \texttt{supplementary\_4.pdf}.

\section{Discussion}
\label{sec:discussion}

This study provides a comprehensive longitudinal analysis of surgeons’ awareness, expectations, and involvement with surgical AI, offering insights into trends within each year (2021 and 2024) while also comparing the two cohorts. It also provides a publicly accessible dataset and codebase for further analysis and reproducibility. The findings reveal a steady increase in awareness of AI tools and participation in AI-related courses, particularly through online platforms, with more than 90\% of surgeons expressing interest in AI education and nearly half advocating for its integration into medical and surgical training. Almost all surgeons believe that AI will eventually be integrated into surgery. Approximately 80\% of respondents anticipate a positive impact, and almost all respondents are willing to adopt AI tools into their clinical workflows (Figure~\ref{fig:results}C). Despite this enthusiasm, ethical concerns, particularly regarding accountability and transparency, remain significant, with more than half of respondents highlighting these as critical barriers. Additionally, while surgeons consistently expressed optimism about AI’s potential to improve clinical practice, notable shifts in expectations were observed over time. Figure~\ref{fig:barriers}B shows that preoperative applications were emphasized in both 2021 and 2024, but in the latter survey, administrative and hospital workflow management also emerged as key areas where AI is anticipated to make a significant impact. By 2024, the focus had shifted further toward clinical translation, with reimbursement strategies becoming a prominent concern alongside validation, trust, and infrastructural challenges, as represented in Figure~\ref{fig:barriers}D. A persistent gap in foundational understanding of AI concepts, such as machine learning and computer vision, was evident in both surveys.

The findings corroborate previous research on AI awareness in surgical practice, which highlighted low familiarity with AI concepts and limited active participation in surgical data science~\cite{36,37}. Notably, over 70\% of surgeons reported difficulty grasping fundamental AI concepts. These trends have also been reported in other medical specialties, such as radiology and endoscopy, further underscoring the interdisciplinary nature of this challenge~\cite{38,39,40}. Despite these barriers, high levels of interest and willingness to adopt AI tools into clinical workflows echo findings from earlier studies~\cite{39,41}.

What sets this study apart is its novel comparison of surgeons’ perceptions before and after the advent of mainstream generative AI models, such as ChatGPT. The study explores how these technological advancements have influenced surgeons’ perspectives, revealing both increased exposure to AI tools and heightened engagement with AI-focused education by 2024 (Figures~\ref{fig:results}A and~\ref{fig:results}B). Although daily exposure to generative AI hasn’t significantly bridged the foundational knowledge gap, it has heightened awareness and interest in integrating AI into medical and surgical practices. This was corroborated by an increase in concerns about reimbursement strategies as surgeons contemplate the clinical translation of AI tools. These findings underscore the importance of incorporating AI-focused curricula into medical training, leveraging surgical associations such as EAES and platforms like Coursera and Stanford Online to enhance accessibility and knowledge dissemination.

The shift in expectations for AI’s role in surgery and the increase in relevance of its applications in administrative work highlights the evolving perception of AI’s utility in clinical settings. This change is consistent with emerging literature suggesting that AI can optimize resource allocation, patient scheduling, and operational efficiency in healthcare systems~\cite{42}. Furthermore, while the potential for integrating AI with AR and VR systems for surgical vision was a key area of interest in 2021, the focus in 2024 shifted toward practical implementation, emphasizing rigorous validation and surgeon control over AI tools. However, enhanced surgical vision with surgical guidance systems was recognized as the most important use of surgical AI. The concept of a master-slave relationship between the surgeon and AI emerged as a central theme, reflecting a strong preference for tools that augment, rather than replace, surgical decision-making.

Concerns regarding trust and accountability also gained prominence in 2024, with surgeons emphasizing the importance of transparent decision-making processes and clear accountability frameworks for AI tools. Collaboration between AI specialists and surgeons was identified as critical for developing systems that integrate within the surgical workflows and address clinical challenges. This interdisciplinary approach is essential not only for building trust but also for addressing persistent barriers such as insufficient data storage infrastructure and limited capabilities for surgical video recording. These challenges, already recognized as significant obstacles in prior studies~\cite{41}, require urgent attention from policymakers and healthcare institutions to enable seamless integration of AI into surgical practice.

This study may suffer some limitations. A selection bias is possible due to the targeted distribution of the surveys, particularly in 2021, when it was shared during the “AI \& Surgery” webinar at IRCAD Strasbourg, drawing participants with a pre-existing interest in AI. In addition, both surveys used non-probability sampling, which limits the generalizability of the findings to the broader surgical community. Lastly, the smaller sample size in 2024 compared to 2021 introduces sampling variability. The demographic distribution of the two surveys, however, does not differ, except for scientific experience. Moreover, except for the venue-specific channels (IRCAD and EAES), dissemination over social media and WebSurg was equally adopted.

\section{Conclusions}
\label{sec:conclusions}

This study highlights a shift in surgeons’ awareness, expectations, and involvement with AI from 2021 to 2024, very likely influenced by the rapid rise of generative AI models. While familiarity with AI concepts remains limited, surgeons are increasingly recognizing the importance of AI in clinical practice and are willing to embrace AI integration, provided that appropriate educational resources, infrastructure, and collaborative development processes are in place. To fully realize AI’s potential in surgery, sustained efforts are needed to promote interdisciplinary collaboration, enhance AI-focused training programs, and address infrastructure and trustworthiness challenges. These steps will ensure that AI contributes positively to surgical practice while maintaining the transparency and accountability that clinicians and patients alike demand.

\section{Acknowledgement}

\subsection{Data Availability Statement}

The data utilized in this study are freely available at the following \href{https://colab.research.google.com/drive/1kQwmi4mdI1W0TYSk-EcdYc3CzNtcgSQ4?usp=sharing}{Google Colab}, and at these Google Sheets files for the raw data: \href{https://docs.google.com/spreadsheets/d/1pvRDJ6Hey4O-FWKNFXnQuDtyaFrBGw24jcKhyDoa9JM/edit?usp=sharing}{2021}, \href{https://docs.google.com/spreadsheets/d/18-pIncaQOd7HSPSQoMKvHlfhl2npoYlVWJ6RklBYQno/edit?usp=sharing}{2024}

\subsection{Funding Statement}

This work was developed within the Interdisciplinary Thematic Institute HealthTech (ITI 2021-2028 program of the University of Strasbourg, CNRS and Inserm),  supported by IdEx Unistra (ANR-10-IDEX-0002) and SFRI (STRAT’US project, ANR-20-SFRI-0012) under the framework of the French Investments for the Future Program.

This work has received funding from the European Union (ERC, CompSURG, 101088553). Views and opinions expressed are however those of the authors only and do not necessarily reflect those of the European Union or the European Research Council. Neither the European Union nor the granting authority can be held responsible for them. This work was also partially supported by French state funds managed by the ANR under Grant ANR-10-IAHU-02.

M.W. was funded by the German Research Foundation (DFG, Deutsche Forschungsgemeinschaft) as part of Germany’s Excellence Strategy – EXC 2050/1 – Project ID 390696704 – Cluster of Excellence “Centre for Tactile Internet with Human-in-the-Loop” (CeTI) of Technische Universität Dresden, and by the Federal Ministry of Education and Research of Germany in the programme of “Souverän. Digital. Vernetzt.”. Joint project 6G-life, project identification number: 16KISK002

\subsection{Declaration of AI in the writing process}

During the preparation of this work, the authors used GPT-4o in order to optimize the writing style. After using this tool/service, the authors reviewed and edited the content as needed and take full responsibility for the content of the publication.

\section*{EAES Working Group}

{\normalsize
Yoav Mintz\textsuperscript{g,h}, Kiyokazu Nakajima\textsuperscript{i}, Michele Diana\textsuperscript{j,k,l}, Tim Horeman\textsuperscript{m}, Manish Chand\textsuperscript{n}, Rosa Maria Jimenez-Rodriguez\textsuperscript{o}, Luigi Manfredi\textsuperscript{p}, Hans Fuchs\textsuperscript{q}, Young Woo Kim\textsuperscript{r}, Martin Wagner\textsuperscript{s,t}, Pieter de Backer\textsuperscript{u}, Felix Nickel\textsuperscript{v}

\vspace{0.8em}
{\small\textit{
\begin{itemize}
  \item[g] Department of General Surgery, Hadassah Hebrew University Medical Center, Jerusalem, Israel
  \item[h] Faculty of Medicine, Hebrew University of Jerusalem, Jerusalem, Israel
  \item[i] Department of Gastroenterological Surgery, Osaka University Graduate School of Medicine, Suita, Japan
  \item[j] University of Strasbourg, CNRS, INSERM, ICube, UMR7357, Strasbourg, France
  \item[k] Department of Surgery, University Hospital of Geneva, Geneva, Switzerland
  \item[l] Medical Faculty, University of Geneva, Geneva, Switzerland
  \item[m] Department of Biomechanical Engineering, Technical University of Delft, Delft, The Netherlands
  \item[n] Division of Surgery and Interventional Science, University College London, London, United Kingdom
  \item[o] University of Seville, Specialist Area Faculty of the Department of General and Digestive Surgery, Specialized Coloproctology Unit, Hospital Universitario Virgen del Rocío, Sevilla, Spain
  \item[p] Division of Imaging Science and Technology, School of Medicine, University of Dundee, Dundee, United Kingdom
  \item[q] Faculty of Medicine and University Hospital of Cologne, Department of General, Visceral and Cancer Surgery, University of Cologne, Cologne, Germany
  \item[r] Center of Gastric Cancer, National Cancer Center, Goyang, Republic of Korea
  \item[s] Department of Visceral, Thoracic and Vascular Surgery, Faculty of Medicine and University Hospital Carl Gustav Carus, TUD Dresden University of Technology, Dresden, Germany
  \item[t] Center for Tactile Internet with Human-in-the-loop (CeTI), TUD Dresden University of Technology, Dresden, Germany
  \item[u] ORSI Academy, Melle, Belgium
  \item[v] Department of General, Visceral, and Transplantation Surgery, Heidelberg University Hospital, Heidelberg, Germany
\end{itemize}
}}

\bibliographystyle{splncs04}
\bibliography{arxiv}
\end{document}